\documentclass[aps,prl,showpacs,twocolumn,groupedaddress]{revtex4-1}

\usepackage[dvips]{graphicx}
\usepackage{dcolumn}
\usepackage{bm}
\usepackage{color}
\def\be{\begin{equation}}
\def\en{\end{equation}}
\def\bea{\begin{eqnarray}}
\def\ena{\end{eqnarray}}
\def\bec{\begin{equation}\begin{array}{rcl}}

\def\p{\partial}
\def\ep{\epsilon}
\def\gs{\gtrsim}
\def\ls{\lesssim}

\newcommand{\av}[1]{\langle{#1}\rangle}

\newcommand{\bi}[1]{\mbox{\boldmath$#1$}}

\def\hrij{\hat{\bi r}_{ij}}

\begin{document}
\title{
Formation of double   glass 
in binary mixtures of anisotropic particles  
}  
\author{Kyohei Takae and Akira Onuki}
\affiliation{Department of Physics, Kyoto University, Kyoto 606-8502, Japan}


\date{\today}

\begin{abstract} 
We study glass transitions in 
mixtures of  elliptic  and circular particles 
in two dimensions using  an orientation-dependent    Lennard-Jones 
potential. 
Changing anisotropic parameters of the potential, the size ratio, and 
the concentration, we realize double glass,  
where both the particle positions and  orientations are   
 disordered but still hold  mesoscopic order. 
The  ellipses  are 
anchored  around the circular impurities  in the  homeotropic 
or planar  directions. 
We examine slowing-down of    rotational and translational 
time-correlation functions. 
Turnover motions of the ellipses are activated  more frequently 
than the  configuration changes, where the latter cause 
the structural relaxation.  
\end{abstract}

\maketitle

Much attention has been 
paid to various types of 
 glass transitions, where  the structural  relaxations become extremely  slow 
with lowering the temperature $T$ \cite{Ang,Binder}. 
In experiments, colloidal particles  can be spherical, but    
 real molecules are mostly nonspherical.  The  
translational and rotational  diffusion constants 
 have  thus   been measured 
in molecular systems 
near the  glass transition \cite{Silescu}. 
Using  generalized  mode-coupling theories,       
some authors    \cite{Rolf,Sch,Gotze} have studied 
the coupled translation-rotation dynamics 
to predict  translational glass and  orientational glass.  
Molecular dynamics simulations 
have also  been performed  on glass-forming binary 
mixtures composed of  anisotropic particles    
   \cite{Klein,Kob1,Chong,Debe,Michele,An}.

For mild differences in  sizes and shapes, 
mixtures of    anisotropic particles 
such as (KCN)$_x$(KBr)$_{1-x}$  form a cubic crystal without 
orientational order (plastic solid) at relatively 
high $T$.  With further lowering  $T$, 
they  undergo a structural phase transition 
in  dilute cases  and become  orientational  
 glass   in nondilute cases   \cite{ori,EPL}. 
On the other hand, if the two species have  
significantly different   sizes or shapes, translational  glass  
 emerges from liquid   at low $T$.  
Here,  the  molecular 
rotations sensitivelly depend on 
the molecular shapes and, for not large 
aspect ratios, flip  motions 
 can occur without large positional displacements. 
In the previous simulations\cite{Klein,Kob1,Chong,Debe,Michele}, 
the  translational and rotational degrees of freedom 
were  strongly coupled.  
Theoretically, for double glass \cite{Sch},  
 they   can be  
simultaneously   arrested   at the same temperature.

In the iterature,  the physical picture of double glass 
remains  quite unclear.  
In this Letter, 
 we    visualize  the particle  
 configurations below  a  double glass transition. 
We shall see marked orientational and positional  heterogeneities 
on mesoscopic scales, where the  latter  
have been detected for circular or spherical particles    
\cite{Shintani,Hamanaka,Tanaka}. 
As a related experiment,  
Zheng \textit{et al.}\cite{China}   
visualized  two-dimensional  motions of  colloidal 
ellipsoids  in monolayers at glass transition.

\textit{Model and numerical method}-  
In two dimensions, we consider   mixtures  
  of  anisotropic  and circular  
 particles with   numbers  $N_1$ and $N_2$, 
where   $N=N_1+N_2=4096$. 
The  concentration  of the circular  species is  
$c=N_2/N.$   The particle   positions are  
written as $\bi{r}_i$ ($i=1,\cdots,N$). 
The orientation vectors of the anisotropic  particles  
 are  expressed as 
$\bi{n}_i=(\cos\theta_i,\sin\theta_i)$
   in terms of angles $\theta_i$  ($i=1,\cdots,N_1$).     
The pair potential $U_{ij}$  between particles 
$i\in\alpha$ and $j\in\beta$ 
($\alpha,\beta=1,2$) is a  truncated modified Lennard-Jones potential. 
If the particle distance $r_{ij}$ is shorter than 
a cut-off $r_c$, it is written as  
\be  
U_{ij}=4\ep\bigg[(1+ A_{ij}) 
\frac{\sigma^{12}_{\alpha\beta}}{r_{ij}^{12}}
-(1+ B_{ij})\frac{\sigma_{\alpha\beta}^6}{r_{ij}^6} \bigg] -C_{ij}. 
\en 
where $\ep$ is the interaction energy. 
We introduce  characteristic lengths  
$\sigma_{1}$ and $\sigma_2$ by setting  
$\sigma_{\alpha\beta}=(\sigma_\alpha+\sigma_\beta)/2$. We set 
  $U_{ij}=0$  for  $r_{ij}>r_c$,  where 
$r_c/\sigma_1$ is 3  for 
$\sigma_{2}/\sigma_1<1$  and is 4.5 for 
$\sigma_{2}/\sigma_1>1$.  
The  $C_{ij}$ ensures  the continuity of $U_{ij}$ 
at $r_{ij}=r_c$.  The particle anisotropy 
is  accounted for  by the 
factors $A_{ij}$ and $B_{ij}$   depending   on 
  the  angles  between   $\bi{n}_i$,  $\bi{n}_j$, and  
the relative direction 
$\hrij=r_{ij}^{-1}(\bi{r}_i-\bi{r}_j)$. We  set  
\bea
A_{ij} &=&  \chi[ \delta_{\alpha 1} 
({\bi n}_i\cdot\hrij)^2+ 
\delta_{\beta 1} ({\bi n}_j\cdot\hrij)^2], \\
B_{ij} &=& \zeta [\delta_{\alpha 1}\delta_{\beta 2}
({\bi n}_i\cdot\hrij)^2+
 \delta_{\alpha 2}\delta_{\beta 1}
 ({\bi n}_j\cdot\hrij)^2], 
\ena
where  $\delta_{\alpha\beta}$ 
is the Kronecker delta,   
 $\chi$ is  the  anisotropy  strength of  
repulsion,  and  $\zeta$ is that of  attraction  between the two species.  
The Newton equations  of motion are  
${m}d^2{\bi{r}}_i/dt^2=-\p{U}/\p{\bi{r}_i}$
 and ${I}d^2{{\theta}}_i/dt^2=-\p{U}/\p{\theta_i}$, where    
  $U=\sum_{i<j}U_{ij}$ is  the  total potential,  
$m$ is  the  mass common to  the two species,  
 and $I$ is the  moment of inertia   of the first species. 
We note that similar angle-dependent  potentials 
were  used for  liquid crystals \cite{Gay,An},  water \cite{water2D}, 
 glass-forming liquids  \cite{Shintani}, and 
 lipids \cite{Leibler}. 

We  treat 
the anisotropic  particles   as ellipses. 
For fixed orientations of two ellipses $i$ and $j$,  
$U_{ij}$  is minimized at 
 $r_{ij}=2^{1/6}(1+A_{ij})^{1/6}\sigma_1$. 
Then varying $\bi{n}_i$ and $\bi{n}_j$ yields 
the  shortest and longest diameters,  
$a_s=2^{1/6}\sigma_{1}$ and $a_\ell=(1+2\chi)^{1/6}2^{1/6}\sigma_{1}$.  
The  aspect ratio   is $a_\ell/a_s=(1+2\chi)^{1/6}$, 
which  is $1.23$ 
for $\chi=1.2$. The ellipses  have  the molecular 
area   $S_1=\pi{a_s}a_\ell/4$ and the  momentum of inertia  
$I=(a_\ell^2+a_s^2)m_1/16.$   
We fix  the average 
 packing fraction    
$(S_1N_1+S_2N_2)/L^2$ at $0.95$,
 where $S_2=\pi2^{1/3}\sigma_{2}^2/4$. 
The system length $L$ is  
about $70\sigma_1$.

We integrated the Newton equations under the periodic boundary 
condition. We  assumed    a Nos\'e-Hoover thermostat \cite{nose}
 to prepare the initial particle 
 configurations, but it was switched off during 
calculating  the time-correlation functions. 
 We measure time  in  units of    
$\tau_0=\sigma_1\sqrt{m/\ep}$ 
and temperature in units of 
$\epsilon/k_B$, where  $k_B$ is   the Boltzmann constant. 
We lowered $T$  from 1 to 0.1 
at  a  cooling rate of  $dT/dt=0.9\times10^{-5}$.
We then changed $T$ to a final temperature 
and waited for $2\times10^5\tau_0$.

At low $T (\ls 0.1)$,  
we observed  only thermal vibrations in 
the particle orientations and positions 
for each  $\chi$,  $\zeta$,   
 $\sigma_2/\sigma_1$, and $c$.    
For not large $\chi$ and $\zeta$,  
we may study  orientational  glass with increasing $c$ 
 \cite{EPL}.  In this Letter, 
we assume  relatively  small or large   $\sigma_2/\sigma_1$  
to induce  disorder  both in the  
orientations and the positions. 
The second species is treated as impurities for $c\ls0.2$.

\begin{figure}
\begin{center}
\includegraphics[width=240pt]{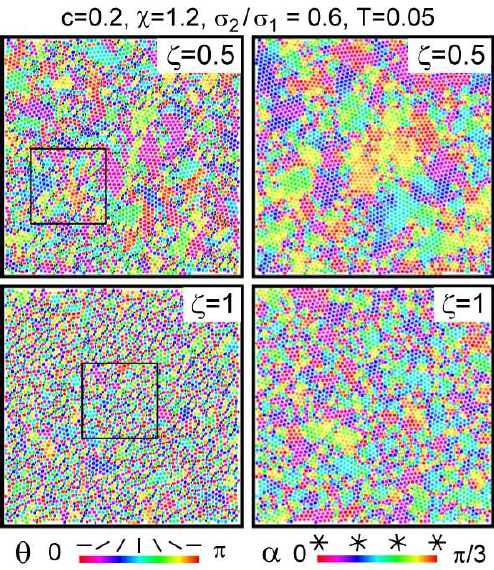}
\caption{Orientational angles 
$\theta_j$ (left) and sixfold bond orientation angles 
$\alpha_j$ (right) in Eq.(3) for small  impurities 
with   $\zeta=0.5$ (top) 
and $1$ (bottom) in double glass, 
where  $c=0.2$,  $\chi=1.2$,   $\sigma_2/\sigma_1=0.6$, 
  and $T=0.05$. Heterogeneities become finer with increasing $\zeta$.}
\end{center}
\end{figure}

\begin{figure}
\begin{center}
\includegraphics[width=240pt]{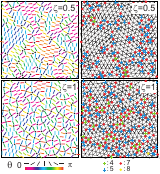}
\caption{Left: Expanded snapshots of orientational angles $\theta_i$ 
around small impurities  in the box regions in the left panels in Fig.1.
Anchoring is homeotropic and  
inpurity clustering is suppressed with increasing $\zeta$. 
Right: Delaunay triangulations, 
where  small impurities 
have  four  or five  surrounding triangles and  
host ellipses have  seven or eight 
triangles.}
\end{center}
\end{figure}

\begin{figure} 
\begin{center}
\includegraphics[width=230pt]{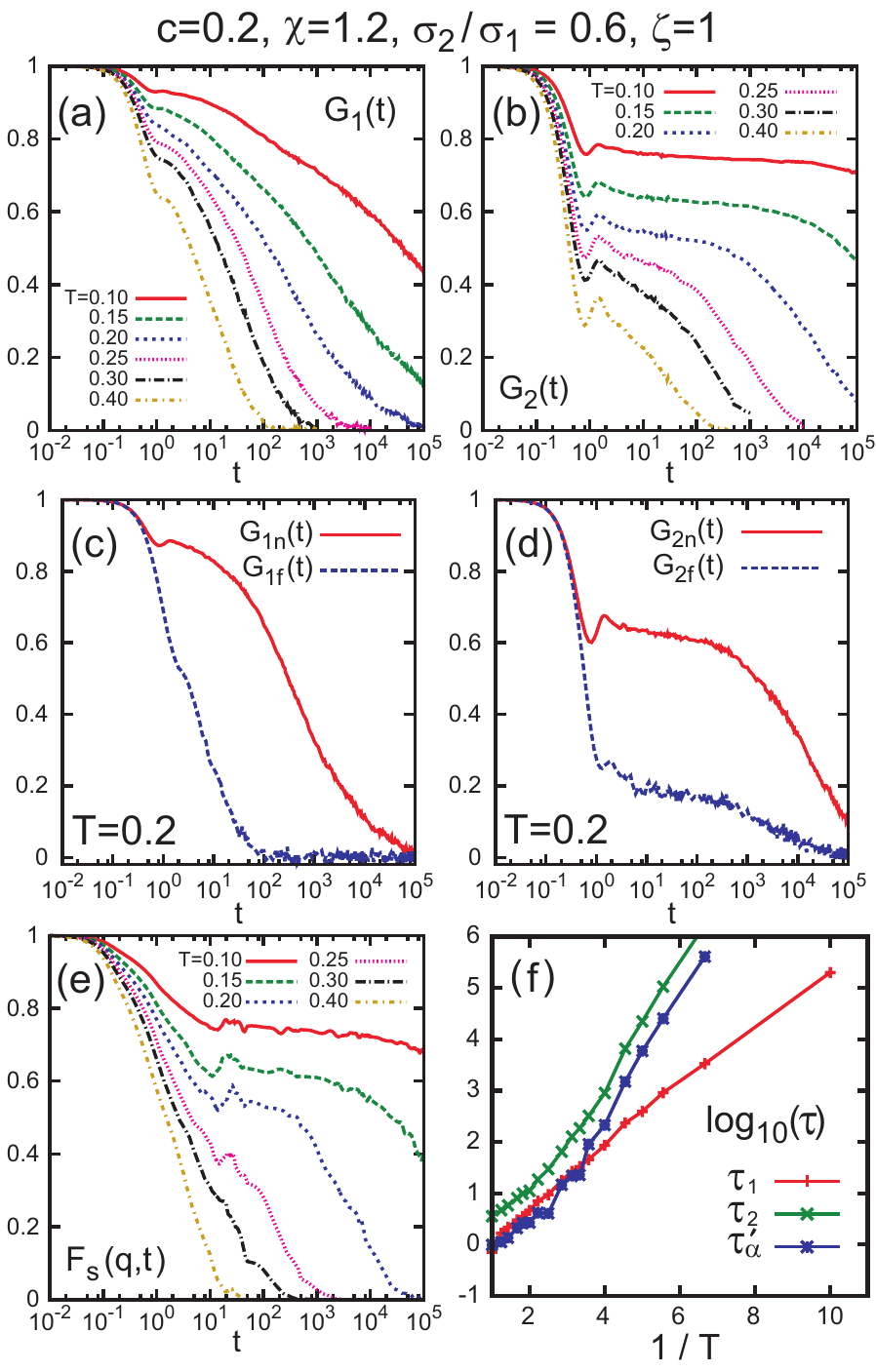}
\caption{ (a) 
$G_1(t)$   and  (b) $G_2(t)$  for six $T$ for ellipses. 
Contributions from ellipses near and far from small  
impurities, $G_{\ell{\rm{n}}}(t)$  and  $G_{\ell{\rm{f}}}(t)$, 
 at $T=0.2$ ($\ell=1$ in (c) and 2 in (d)). 
 $F_{s}(q,t)$  at $q=2\pi$ in (e) 
and   relaxation times  $\tau_1$, $\tau_2$, and $\tau_{\alpha}'$ vs $1/T$ 
in (f). As in the bottom  in Fig.1, 
 $c=0.2$, $\chi=1.2$, $\sigma_2/\sigma_1=0.6$, and $\zeta=1$. 
Time $t$ is  in units of 
$\sigma_1(m/\epsilon)^{1/2}$.}
\end{center}
\end{figure}

\textit{Small impurities}-  
First, we consider  
 small impurities with  $\sigma_2/\sigma_1=0.6$ 
 setting   $\chi=1.2$,   and  $c=0.2$. 
We are interested in  the effect of   the anisotropic   attraction  arising 
from  $\zeta>0$. At  $T=0.05$,   
 Fig.1 displays 
  the angles $\theta_j$ of  the ellipses  
and the sixfold bond-orientation  angles $\alpha_j$  
for all the particles. In the range $0\le\alpha_j<\pi/3$, 
we define  $\alpha_j$ by   \cite{Nelson,Hamanaka} 
\be 
\sum_{k\in\textrm{\scriptsize{bonded}}} \exp [6i\theta_{jk}]
\propto  \exp[{6i\alpha_j}], 
\en 
where    $\theta_{jk}$ 
is the angle between $\bi{r}_k-\bi{r}_j$ and  the $x$ axis, 
 the bonded particles $k$ are  within the range 
 $|\bi{r}_{jk}|<1.5\sigma_{\alpha\beta}$, 
and $6\alpha_j$ is the   phase angle   of the left hand side. 
 
In Fig.1,   orientationally  ordered regions 
are finely divided  by the impurities.   
For $\zeta=0.5$ we can see  small 
orientationally ordered domains and small polycrystal grains, 
while for  $\zeta=1$  orientational   and positional   
disorders are both  enhanced,  resulting in  double glass.   
Namely, with increasing $\zeta$,  the orientation domains 
and the grain sizes become smaller. 
Similar positional patterns 
have been observed  in glass \cite{Hamanaka,Shintani,Tanaka}. 
Note that the crossover between polycrystal and glass 
is gradual and mesoscopic order still remains in glass.

The left panels of Fig.2 display  expanded snapshots 
of $\theta_j$ around impurities,  where anchoring is 
homeotropic  \cite{An}.   With increasing $\zeta$,  
the  ellipses tend to  attach 
 to each impurity and the  impurity clustering 
is gradually   suppressed, where 
the clustering  took place during solidification.  
Similar  homeotropic anchoring occurs  in water  around  
small ions as hydration, which breaks   tetrahedral order 
resulting in   vitrification  at low $T$ 
\cite{water}. 

In the right panels of Fig.2, we show the Delaunay triangulation 
of the same positional  configurations. 
 Here, the number of surrounding triangles $n_{\rm t}$ 
deviates from six. For small impurities we have  
$n_t =4$ or 5 mostly and , if $\zeta=0.5$ ($\zeta=1$), their 
fractions  are $5\%$ ($29\%$) for $n_{\rm t} =4$  
 and   $94\%$ ($71\%$) for $n_{\rm t} =5$.  
For large host particles 
we have $n_t =6$ or 7 mostly and, if  $\zeta=0.5$ ($\zeta=1$), their 
fractions  are   $73\%$ ($66\%$) for $n_{\rm t} =6$ 
 and   $26\%$ ($33\%$) for $n_{\rm t} =7$. 
In liquid,   small  host  particles 
noticeably  form 
liquidlike defects with   $n_t>6$ ($2.5\%$ at $T=0.4$).   
 Hentschel {\it et al.} \cite{Pro} 
used  Voronoi graphs (dual to  Delaunay ones) 
for a mixture of circular particles. In accord with our result, 
they found  that 
 most small (large) particles are enclosed in  pentagons 
(heptagons)  in glass. 
In their theory, this is a characteristic feature 
of translational  glass.

In Fig.3(a)-(b), we  plot  
 the rotational time-correlation functions 
$G_1(t)$ and $G_2(t)$ \cite{Klein,Michele,Kob1,Chong,Debe} defined by   
\be 
G_\ell(t)
=\frac{1}{N_1}\sum_{1\le j\le N_1}
 \av{\cos[\ell \theta_j(t+t_0)-\ell \theta_j(t_0)]} ,   
\en 
where $\ell=1,2$. We took the  average $\av{\cdots}$  over 
the initial time $t_0$ and over five  runs.  
Here,   $G_1(t)$ decays    due to 
turnover motions $\theta_j\to\theta_j\pm\pi$, 
while   $G_2(t)$ is unchanged by them. Such flips  
 have been observed in some  
simulations \cite{Kob1,Klein,Michele}. 
While  $G_2(t)$ exhibits a considerable 
initial decay, it  decays more slowly for $t\gs1$. 
In Fig.3(c)-(d), we divide the ellipses $j\in1$  
into those  near  the impurities 
($r_{jk}<1.8\sigma_1$ for some $k\in2$) 
and those far from  the impurities  
($r_{jk}>1.8\sigma_1$ for any  $k\in2$).  
Their numbers are given by 
$N_{\rm{n}}\sim  2400$ 
and $N_{\rm{f}}\sim  900$  and 
their  rotational time-correlation functions 
are written as $G_{\ell{\rm{n}}}(t)$   
and  $G_{\ell{\rm{f}}}(t)$, respectively. 
In our case,   the  impurities strongly  
 anchor  the nearby  ellipses, so the  decay of 
$G_{\ell{\rm{n}}}(t)$  is much slower than that of 
 $G_{\ell{\rm{f}}}(t)$. On the other hand, 
Fig.3(e)  gives  
 the self part of the density time-correlation 
functions $F_{s}(q,t)$ at $q=2\pi/\sigma_1$ for the 
first species. Notice close resemblance between $G_2(t)$ and 
$F_{s}(q,t)$.  In Fig.3(f), we plot 
 the corresponding  relaxation times 
$\tau_1$,  $\tau_2$, and 
${\tau_{\alpha}}'$, where   ${\tau_\alpha}'\cong\tau_2\gg\tau_1$. 
 They are determined by 
$G_1(\tau_1)=1/e$, 
 $G_2(t)\propto\exp[-(t/\tau_2)^{\beta}]$ for  $t>1$, 
and $F_{s}(q,t)\propto\exp[-(t/{\tau_{\alpha}}')^{\gamma}]$ for $t>1$,  
where the  exponents $\beta$ and $\gamma$ 
 are about 0.4  for $T\sim0.2$.  Here,   $\tau_1$ exhibits 
the  Arrhenius behavior $\ln\tau_1\propto1/T$. 
The turnover motions are thermally activated  
 and are decoupled 
from the translational motions. However, 
the slow decay of $G_2(t)$ and 
the positional  relaxation are strongly coupled,  
as a characteristic feature of double glass 
 \cite{Chong,Silescu,Sch,Gotze,Rolf}.

For $T \gs 0.2$, we could realize  steady states. 
However, for $T\ls  0.15$, $\tau_\alpha (\sim \tau_2)$  exceeds $ 10^5$ 
and the waiting time 
was  too short and the 
aging process continued   in the whole 
simulation time  ($\sim 10^6$). 
At $T=0.05$ in Figs.1 and 4,
 there was no  appreciable configuration 
changes in the orientations and positions 
in our  simulation time."

\begin{figure}[htbp]
\begin{center}
\includegraphics[width=240pt]{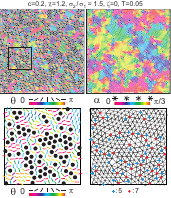}
\caption{Top: 
Orientational angles 
$\theta_j$ (left) and sixfold bond orientation angles 
$\alpha_j$ (right)  for large impurities,  
where  $c=0.2$,   $\chi=1.2$, $\zeta=0$, $\sigma_2/\sigma_1=1.5$, 
  and $T=0.05$. Here the orientational disorder is 
more enhanced than the translational one. 
Bottom: Orientational angles $\theta_i$ 
around  impurities in planar alignment 
(left) and Delaunay triangulation 
exhibiting point defects (right). 
These are expanded snapshots of 
 the box region  in the top left panel.   
}
\end{center}
\end{figure}

\begin{figure}
\begin{center}
\includegraphics[width=230pt]{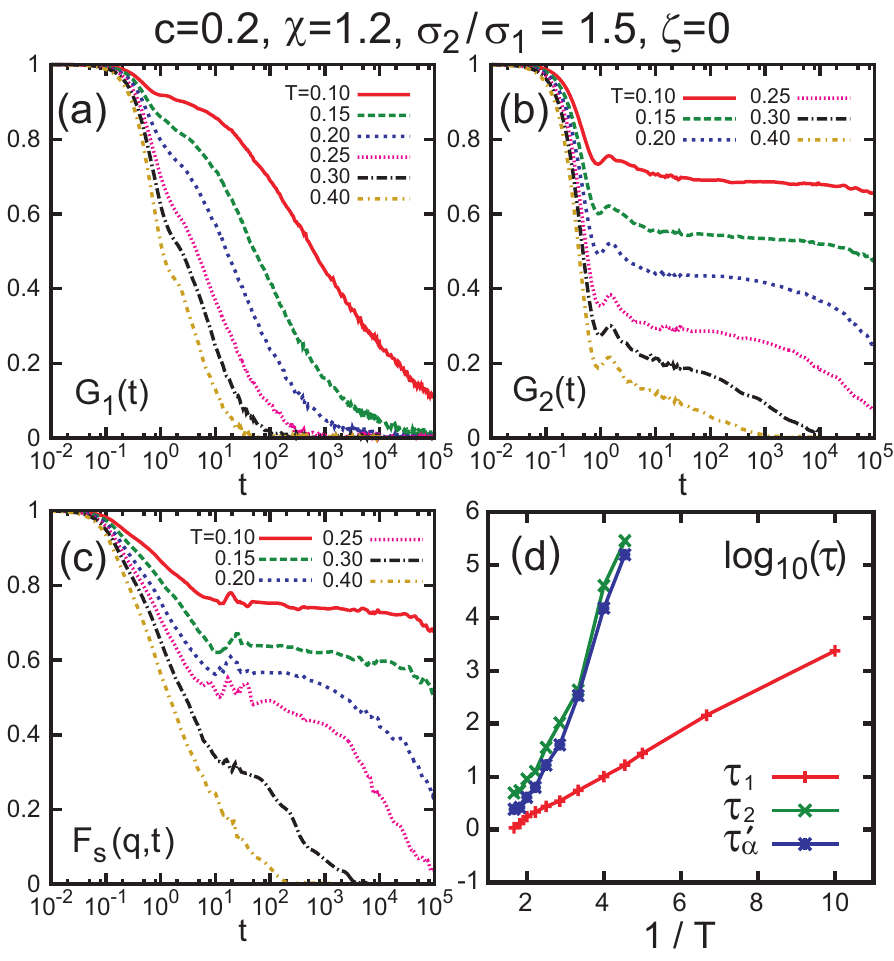}
\caption{ (a) $G_1(t)$   and  (b) $G_2(t)$  for six $T$ for ellipses. 
 $F_{s}(q,t)$  at $q=2\pi$ in (c) and  relaxation times for 
  at $q=2\pi/\sigma_1$ in (d).  Here impurities  are large 
 with $\sigma_2/\sigma_1=1.5$. 
The  parameters are the same as  in  Fig.4. 
Time $t$ is in units of $\sigma_1(m/\epsilon)^{1/2}$.}
\end{center}
\end{figure}

\textit{Large impurities}- 
Next, we consider   
  large impurities with  $\sigma_2/\sigma_1=1.5$  setting   
$\chi=1.2$,  $\zeta=0$, and   $T=0.05$. 
In the upper plates in Fig.4, the orientations are highly frustrated, 
while the particles  still form polycrystal with very small sizes.     
In the lower plates,    we show that the anchoring is 
planar  at the impurity surfaces   \cite{An} (left).  
In this case, impurity clustering is conspicuous, because 
association of impurities  lowers the total potential energy 
by a few $\epsilon$ per impurity particle  \cite{EPL}. 
We also show  point defects in the Delaunay triangulation (right). 
Here,   we have  
$n_{\rm t} =6$ ($67\%)$  or 7 ($33\%)$ for large  impurities 
and  $n_{\rm t} = 6$ ($89\%$) or $ 5$ ($10\%)$ 
for small  host particles. In liquid, host 
small particles form  
appreciable liquidlike defects ($2\%$ at $T=0.4$).

In Fig.5, we show the rotational time correlation functions 
$G_1(t)$ and  $G_2(t)$ in (a) and (b), the  
  self part of the density time-correlation function 
$F_s(q,t)$ at $q=2\pi$ in (c), 
and relaxation times 
$\tau_1$, $\tau_2$, and ${\tau_\alpha}'$ vs $1/T$ in (d). 
These quantities are defined  in Fig.3.    
Their  behaviors, including those of 
  $G_{\ell{\rm{n}}}(t)$   
and  $G_{\ell{\rm{f}}}(t)$, 
 are also very similar to those in Fig.3, though 
the types of impurities are very different in the two cases. 
We again find ${\tau_\alpha}'\sim\tau_2\gg\tau_1$ and 
the Arrhenius behavior of    $\tau_1$.

\textit{Summary and remarks}- With  an   angle-dependent 
Lennard-Jones potential, we have performed 
simulation of  mixtures of elliptic and circular 
particles. For mild anisotropy, 
the  ellipses  tend to form a lattice of 
isosceles triangles  far from the impurities but   
 are anchored  around them  in  
 homeotropic or planar alignment at low $T$. 
The positional disorder is produced 
if the size ratio $\sigma_2/\sigma_1$ 
considerably  deviates from unity. In such cases,  
the  orientational order 
and the positional order   decrease 
with increasing the impurity concentration $c$. 
For small impurities, their clustering is 
suppressed with increasing the 
 impurity-ellipse interaction ($\propto \zeta$) in Fig.2. 
For large impurities, it can also be 
suppressed with increasing repulsion among them, for example, 
by adding a term 
proportional to $\delta_{\alpha 2}\delta_{\beta 2}$ 
in $A_{ij} $ in Eq.(2). 

We  have also  studied the rotational dynamics 
of the ellipses, whose  
 turnover motions occur  more frequently than  
  the configuration changes in agreement with the previous papers. 
  In addition, 
the ellipses near the impurities 
 rotate more slowly 
than  those far from them.

Our  potential energy is invariant with respect to the  turnover motions. 
However, if it has no  such flip symmetry, 
a flip-translation coupling arises. 
For example, $\tau_1$ increases  
with introduction of the dipolar interaction, 
which will be reported shortly.

The spatial scales of the    structural 
 heterogeneities  
 depend on various parameters as in Figs.1 and 4. 
If the  oriented domains are not too small,   
there arises a  large orientation-strain coupling, 
leading to   soft elasticity and 
 a shape-memory effect  \cite{EPL}. Such effects were  found  
for  Ti-Ni  alloys \cite{Ren} (where atomic displacements 
within unit cells cause structural   changes). 
When anisotropic particles  have electric dipoles\cite{ori}, 
we will report appearance of mesoscopic  polar  domains  
yielding  large response to  electric field,    
as in ferroelectric relaxors \cite{Cowley}.

\begin{acknowledgments}
This work was supported by Grant-in-Aid 
for Scientific Research  from the Ministry of Education, 
Culture,  Sports, Science and Technology of Japan. 
K. T. was supported by the Japan Society for Promotion of Science.
\end{acknowledgments}

\end{document}